\documentclass[a4paper]{jpconf}
\usepackage{graphicx}
\usepackage{epstopdf}
\usepackage{bm}
\usepackage{comment}

\newcommand{\beq}{\begin{equation}}
\newcommand{\eeq}{\end{equation}}
\newcommand{\bea}{\begin{eqnarray}}
\newcommand{\eea}{\end{eqnarray}}

\def\bx{\bf x}

\def\OMIT#1{{}}

\newcount\hour \newcount\hourminute \newcount\minute 
\hour=\time \divide \hour by 60
\hourminute=\hour \multiply \hourminute by 60
\minute=\time \advance \minute by -\hourminute

\begin{document}
\title{Lattice QCD at non-zero isospin chemical potential}

\author{Zhifeng Shi$^{1, 2}$}

\address{$^1$Department of Physics, The College of William \& Mary, Williamsburg, Virginia, USA}
\address{$^2$Jefferson Lab, Newport News, Virginia, USA}

\ead{ zshi@email.wm.edu }

\begin{abstract}
 Systems of non-zero isospin chemical potential
 are studied from a canonical approach by computing
 correlation functions with the quantum numbers of 
 $N\ \pi^+$'s ($C_{N \pi}$). In order to reduce the number of contractions
 required in calculating $C_{N \pi}$ for a large $N$
 in the Wick's theorem, we constructed a few new
 algorithms. With these new algorithms, systems with 
 isospin charge up to $72$ are investigated on 
 three anisotropic gauge ensembles with a pion mass of $390\ \rm{MeV}$,
 and with lattice spatial extents $L \sim {2.0,\ 2.5,\ 3.0}\ \rm{fm}$.
 The largest isospin density of $\rho_I \approx 9\ \rm{fm}^{-3}$ is 
 achieved in the smallest volume, and the QCD phase diagram
 is investigated at a fixed low temperature at varying isospin chemical
 potentials, $m_{\pi} \le \mu_I \le 4.5\ m_{\pi}$. By investigating 
 the behaviour of the extracted energy density of the system
 at different isospin chemical potentials,
 we numerically identified the conjectured transition to a Bose-Einstein condensation
 state at $\mu_I \ge m_{\pi}$.
\end{abstract}

\section{Introduction}
An important goal of nuclear physics is to investigate 
interactions between hadrons;
 however studying many body systems 
are hampered by the factorially growing number of Wick contractions
naively required in computing the corresponding correlation function.
In order to study multi-meson systems, a few algorithms~\cite{Detmold:Savage,Detmold:Kostas:Zhifeng_2012} have been constructed
to alleviate the cost of computing larger number of independent contractions.
With these new algorithms,
many-pion systems have been studied in 
Ref.~\cite{Detmold:Kostas:Zhifeng_2012,Beane:2007es,Detmold:2008fn},
and many-kaon systems have been studied in Ref.~\cite{Detmold:2008yn},
furthermore systems of 
mixed species have also been studied in Ref.~\cite{Detmold_Brian:2011}.
In order to study more complicated 
multi-baryon systems,
various methods have also been introduced.
With methods constructed in \cite{Detmold:Kostas:_2012},
correlation functions of $^{28}$Si have been computed within a manageable 
amount of time.

Knowledge of  systems at 
high density, with non-zero chemical potential, and non-zero temperature
are vital to explore the QCD phase diagram.
Systems with zero chemical potential have been investigated
for a wide range of temperatures, and the transition from confined 
state at low temperature to the de-confined state at high
temperature has been investigated substantially~\cite{c.detar_2012}. 
However, direct simulations of non-zero chemical potential systems 
are much harder due to the famous sign problem
resulting from the non positive fermion determinants.
But the sign problem does not exist for 
isospin chemical potential systems. Although QCD phase diagrams of 
these two systems are not the same, they do share some common properties,
and it is also theoretical interesting to study non-zero isospin chemical potential
systems.

In this proceeding,
systems of non-zero isospin chemical potential are studied from a canonical 
approach by directly computing correlation functions of $n$-$\pi^+$ systems.
By appling algorithms constructed in \cite{Detmold:Kostas:Zhifeng_2012},
we studied systems containing up to $72\ \pi^+$'s on 
three anisotropic $2+1$ flavor dynamical gauge ensembles, $16^3 \times 128$,
$20^3 \times 256$ and $24^3 \times 128$, with $\xi = a_s/a_t = 3.5$,
$a_s = 0.125$ fm, and with a poin mass of $m_{\pi} = 390$ MeV and a kaon mass
of $m_{K} = 540$ MeV.
The QCD phase diagram at a fixed low temperature, ${\cal{T}}\sim 20$ MeV, for
a range of isospin chemical potentials, $m_{\pi} \le \mu_I < 4.5\ m_{\pi}$ ,
 are investigated, and evidence for a phase transition
from pion gasses to Bose-Einsetin Condensate (BEC) state around $\mu_I = 1.3 m_{\pi}$
is numerically identified. 

The layout of this proceeding is as follows. In Sec.\ref{sec:method}, we briefly 
discuss the method applied to study $n$-$\pi$ systems. Main results from the 
simulation are presented in Sec.\ref{sec:results}, and we conclude in Sec.\ref{sec:conclusion}.

\section{Methodology}
\label{sec:method}
Non-zero isospin chemical systems can be investigated from the study of 
$n$-$\pi$ systems by directly computing their
correlation functions.  Because of the Pauli principle,
the largest number of pions can be studied from a single source location is
$N_s N_c=12$. In order to study systems of more than $12$ $\pi$'s, 
additional source locations
are required. The correlation function for a system of total
\mbox{$\overline{n}=\sum_{i=1}^Nn_i$ $\pi^+$'s} with $n_i\ \pi^+$'s from
source locations (${\bf y}_i,0$) is defined as:
\begin{eqnarray}
  C_{n_1, ... ,n_N}(t) & = &
  \left <\
    \left(\ \sum_{\bf x}\ \pi^+({\bf x},t)\ \right)^{\overline{n}}
    \left( \phantom{\sum_{\bf x}}\hskip -0.2in
      \pi^-({\bf y_1},0)\ \right)^{n_1} \ldots
    \left( \phantom{\sum_{\bf x}}\hskip -0.2in
      \pi^-({\bf y_N},0)\ \right)^{n_N}\
  \right >
  \ ,
  \label{eq:mpi}
\end{eqnarray}
where the interpolating operator ${\pi^+({\bx},t)} = {\overline
  d({\bx},t)} \gamma_5 u({\bx},t)$ and ${\pi^-({\bx},t)} = {\overline
  u({\bx},t)} \gamma_5 d({\bx},t)$. 
  
 Computing the above correlation function
 includes ${\cal O}({\overline n}!)$ contractions. For larger $\overline n$ and $N$, 
 number of ways to distribute pions over $N$ sources 
 grows exponentially, which further increase the computational
 cost of all $C_{n_1,\cdots, n_N}$'s. However, correlation functions 
 of $\overline n$ pions, $C_{\overline n}$, without knowing details of the 
 distributions of $n_i$ can 
 be constructed, which significantly reduces the number of correlation
 functions to compute as energies extracted from $C_{\overline n}$ is same as those
 extracted from $C_{n_1,\cdots, n_N}$'s as long as $\overline{n}=\sum_{i=1}^Nn_i$.
  The explicit methods constructed in Ref.\cite{Detmold:Kostas:Zhifeng_2012}
 are used to calculate $C_{\overline n}$
  in this proceeding. Detailed discussions of the method and evaluations of $n$-$\pi$
  correlation functions in momentum space can be found in Ref.\cite{Detmold:Kostas:Zhifeng_2012}.

\section{Lattice Results}
\label{sec:results}
Computations are performed on ensembles of 
 $n_f=2+1$ anisotropic gauge configurations with clover-improved fermions generated by
the Hadron Spectrum Collaboration and the Nuclear physics with
Lattice QCD collaboration. 
Pion correlations function, $C_{{\bar n}\pi}(t)$, for ${\bar n}=1, 2, \cdots, 72$,
are computed from three lattice ensembles of $L^3 \times T = 16^3 \times 128,
20^3 \times 256, 24^3 \times 128$. More details of these gauge ensembles 
can be found in Ref.\cite{shei_wohl_83}.

The expected form of $C_{n\pi}(t)$
computed on a lattice
 with temporal extent $T$ is ~\cite{Detmold_Brian:2011}
\begin{eqnarray}
  \label{eqn:corr_expected}
  C_{n\pi}(t) &=& \sum_{m=0}^{\lfloor \frac{n}{2} \rfloor} 
     {n \choose m} A_m^n Z_m^n e^{-(E_{n-m}+E_{m})T/2}
  \cosh((E_{n-m}-E_{m}) (t-T/2))+\ldots, \ \ \ \ 
  \label{eq:c_all_thermal}
\end{eqnarray}
where $A_m^n = 1$ when $m = n/2$, otherwise $A_m^n=2$.
$E_{m}$ is the ground state energy of a $m$-$\pi$ system,
the $Z_m^n$ are overlap factors of thermal states 
with $m\ \pi$'s
propagating backward around the temporal boundary
\footnote{When $m=0$, $Z_0^n$ denotes the ground state 
contribution.}, and
the ellipsis denotes contributions from excited states. 
For infinite $T$, all thermal states vanish, and only
the ground state survives, however the ground 
state is always contaminated by thermal states
for finite $T$.
Even for $C_{12\pi}$ computed on $T=128$
ensembles, the ground state does not dominate in any region.
Under the current precision, $C_{n\pi}$ computed  from ensembles 
with a temporal extent $T=256$ shows a clear plateau region even for $n=72$.
Thus, in order to reduce thermal contaminations, the 
anti-periodic plus periodic propagation ($A \pm P$) method
~\cite{Detmold:Kostas:Zhifeng_2012} has
been applied to effectively double the temporal extent of $T=128$ ensembles. 

\subsection{Ground state energies}
\begin{figure}
  \includegraphics[width=8.5cm]{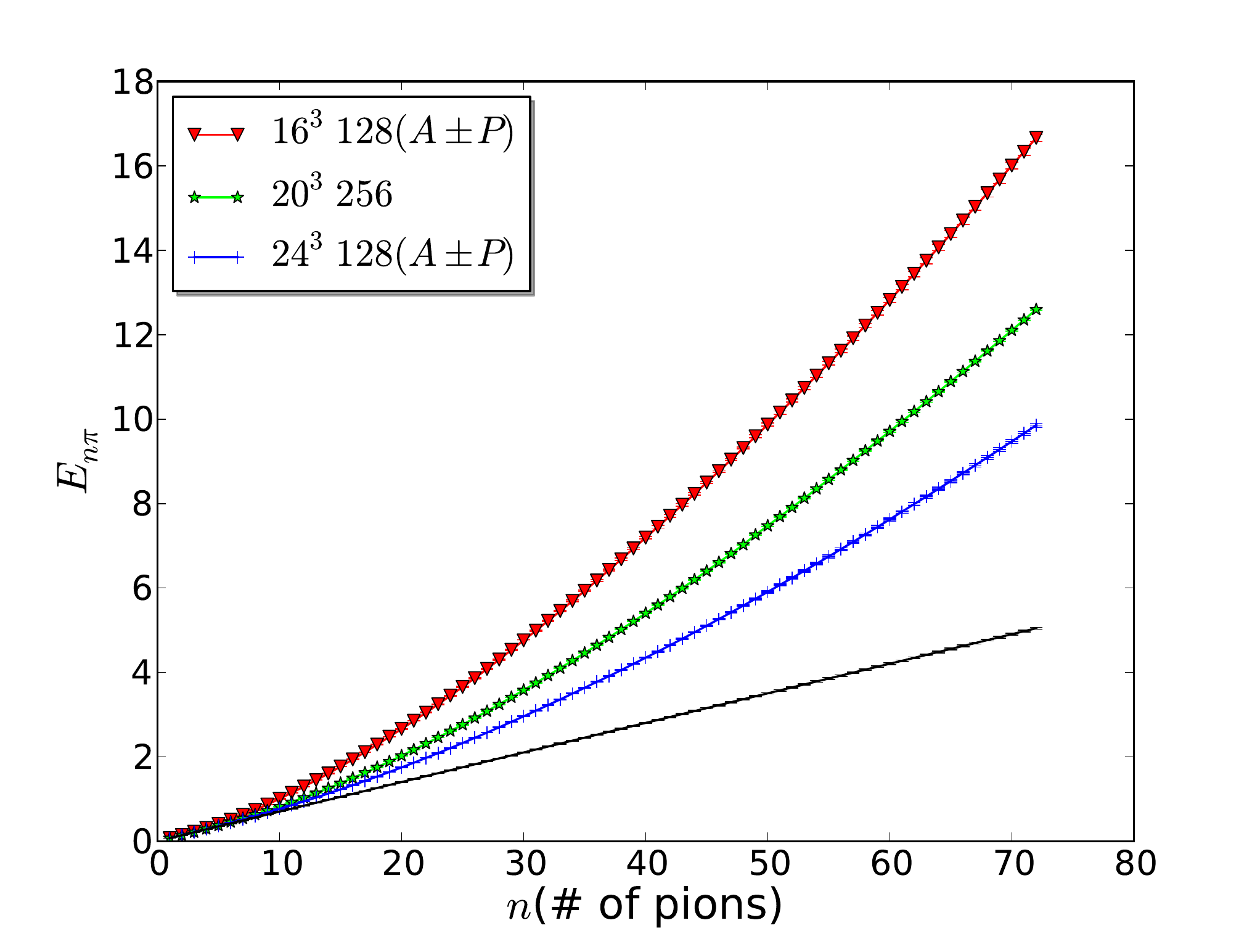}
  \includegraphics[width=8.5cm]{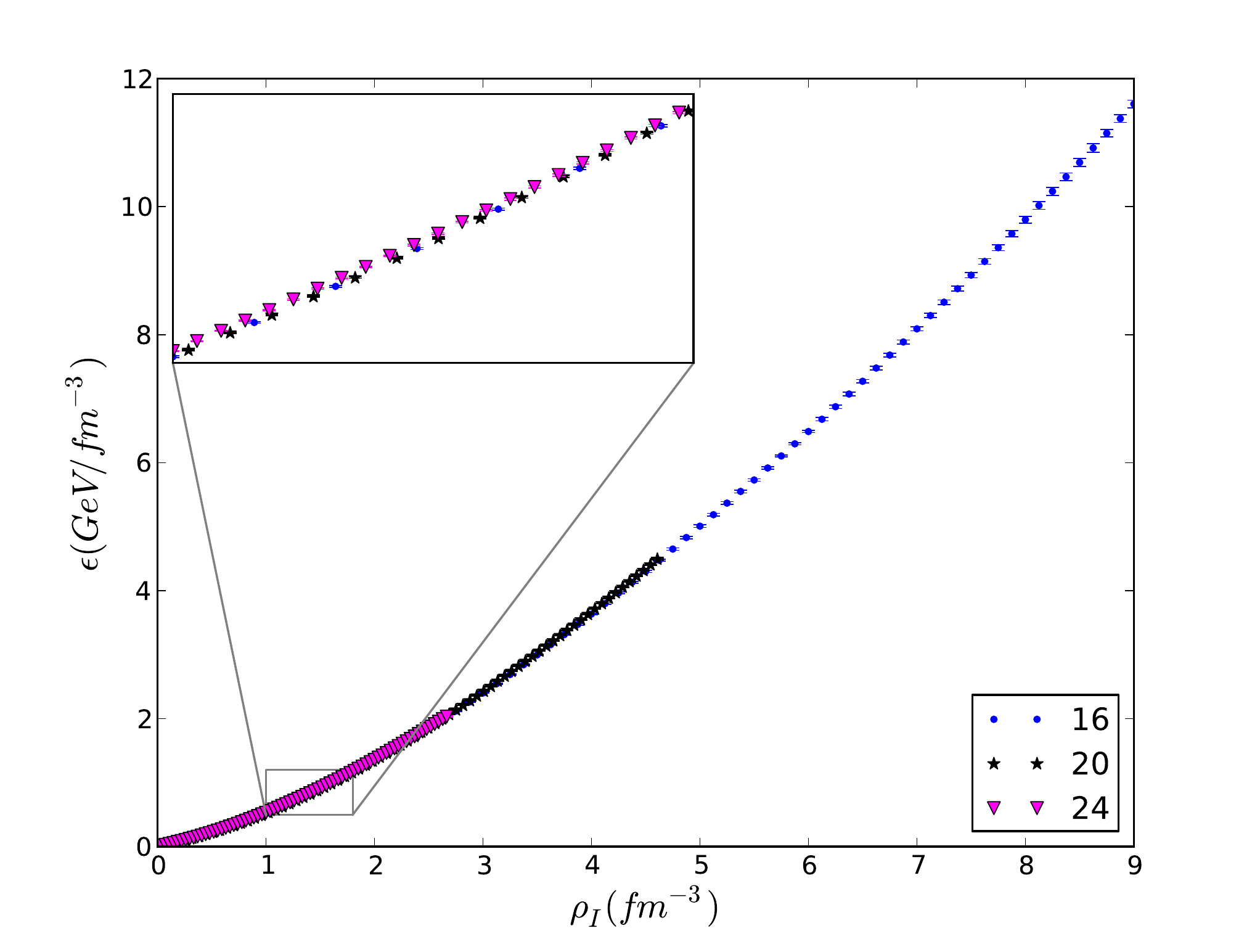}
  \caption{ On the left plot, ground state energies of 
    $n$-$\pi$($E_{n\pi}$) systems extracted from ensembles $16^3 \times 128$ (red), 
    $24^3 \times 128$
    (blue) and $20^3 \times 256$ (green) are shown.  The black line represents the
    free energy of $n$ non-interacting pions.
    Energy densities, $\epsilon(\rho_I)$, computed from three ensemble are compared
    on the right plot.}
  \label{fig:ground_state_energy}
\end{figure}

Ground state energies of $n$-$\pi$ systems are extracted by fitting 
a single exponential to $C_{n\pi}(t)$
within time slices where a clear plateau can be identified
in the corresponding effective mass plot.
Statistical uncertainties are evaluated from the bootstrap
method, and systematic uncertainties are from averaging systematic uncertainties
computed
on each bootstrap sample by moving the fitting window forward and backward 
two time slices. Extracted ground state energies from three volumes are shown 
in the left plot of Fig.~\ref{fig:ground_state_energy}, and the corresponding
energy densities, $\epsilon=E/V$,  are compared in the right.
 Because of the repulsive interactions
between pions, $E_{n\pi}$ is larger in smaller volumes as pions are closer to 
each other, however energy densities computed from three volumes
are approximately the same.

\subsection{Isospin chemical potential}
\begin{figure}
  \includegraphics[width=15.0cm]{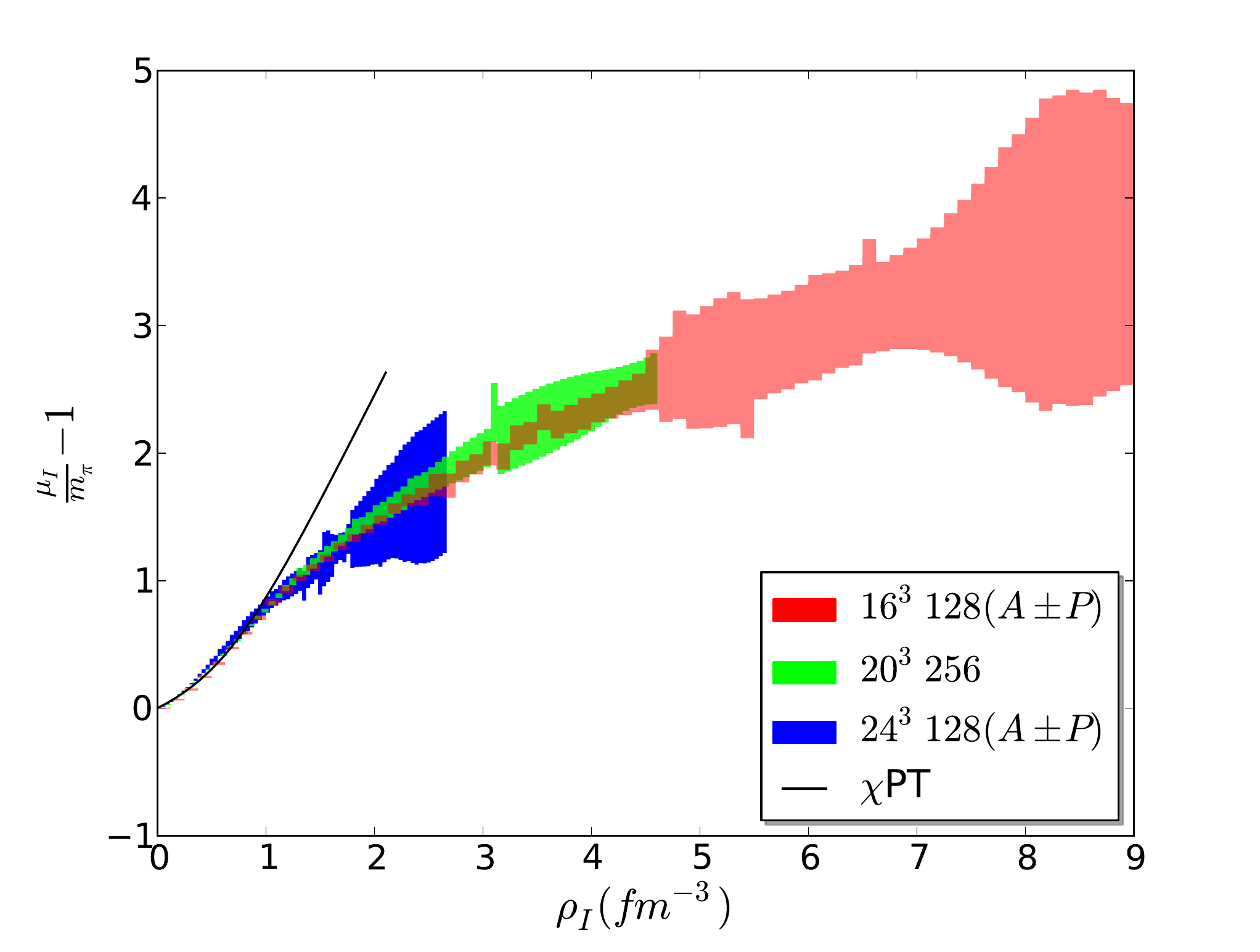}
  \caption{The isospin chemical potential, $\mu_I$, is plotted as a
    function of the isospin density, $\rho_I$, from three lattice
    ensembles, $16^3 \times 128$ (red, $\rho_I = [0,9]$),
    $20^3 \times 256$ (green, $\rho_I = [0,4.7]$) 
    and $24^3 \times 128$ (Blue, $\rho_I = [0,2.8]$).  The solid
    black line is from $\chi$-PT~\cite{Son:Stephanov} }
  \label{fig:chemical_potential}
\end{figure}

From the ground state energies of $n$-$\pi$ systems ($E_{n\pi}$), 
an effective 
isospin chemical potential $\mu_I(n) = \frac{d E }{d n}$ can be computed from
a finite backward derivative, $\mu_I(n) = E_{n\pi}-E_{(n-1)\pi}$. 
 In order to take into account the correlation between
 $E_{n\pi}$ and $E_{(n-1)\pi}$ extracted from the same 
 bootstrap sample, $\mu_I(n)$ is 
also evaluated on each bootstrap sample. The extracted isospin chemical potential 
as a function of isospin density is plotted in Fig.~\ref{fig:chemical_potential}.

At small isospin density, $\mu_I$ grows in an accelerating rate, 
agreeing with the prediction from $\chi$-PT~\cite{Son:Stephanov},
 however it starts to deviate 
from the $\chi$-PT and begins to grow in a decelerating rate at large
isospin densities, and at even larger isospin density it begins to 
flatten out.

\subsection{QCD phase diagram}
\begin{figure}
  \includegraphics[width=15cm]{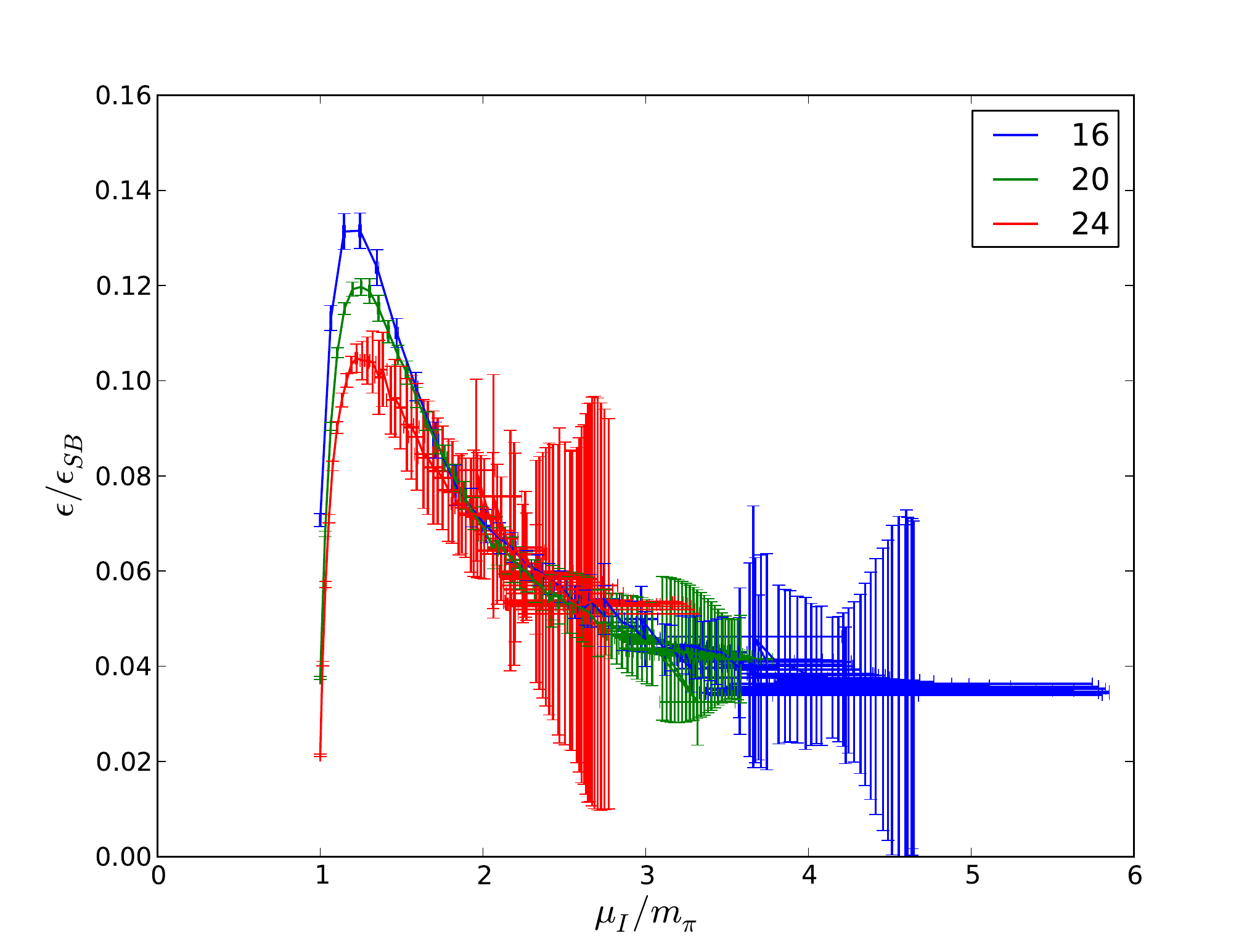}
  \caption{The $\epsilon/\epsilon_{SB}$ is plotted as a function of
    $\mu_I/m_\pi$.}
  \label{fig:epsilon_ratio}
\end{figure}

The change of the behavior of the isospin chemical potential at
different isospin density signals a possible change of the physical state 
of the $n$-$\pi$ system.
In order to investigate this change, we have studied the ratio of
the energy density, $\epsilon$, to its 
zero temperature Stefan-Boltzmann limit,
\begin{eqnarray}
  \epsilon_{SB} = \frac{N_f N_c}{4 \pi^2}\mu_I^4,
\end{eqnarray}
where $N_f = 3$ and $N_c = 3$ is used in this formula.
The ratio is plotted in Fig.~{\ref{fig:epsilon_ratio}}, where
it grows at small 
$\mu_I$, reaches a peak around $1.3\ m_{\pi}$,
starts to decrease after the peak, and begins to flatten out 
at large isospin chemical potential. With a linear extrapolation 
of the peak position for different volumes as a function of the inverse
of volume, the peak position in the infinite volume limit 
is $\mu_{peak}^{I} = 1.30(7)\ m_{\pi}$.

At low temperature, as soon as $\mu_I$ reaches $m_{\pi}$, one pion
is produced out of vacuum, and the system is an
interacting pion gas. When $\mu_I$ reaches the peak at $\mu_I = 1.30(7)\ m_{\pi}$,
pions start to condense and the $n$-$\pi$ system becomes
 Bose-Einsten Condenstate (BEC). Such a  transition to BEC state at $\mu_I > m_{\pi}$
 are also observed in two flavor QCD~\cite{Kogut:2004zg}, and similar behavior occurs in
 two-color QCD~\cite{simon:seyong}.

\section{Conclusion}
\label{sec:conclusion}

In this proceeding, we studied $n$-$\pi$ systems from a canonical
approach by explicitly computing 
correlation function of the $n$-$\pi$ system, $C_{n}(t)$. Ground state energies
of the $n$-$\pi$ systems were extracted, and subsequently, the
isospin chemical potential was computed as a function of the isospin density.

By studying the ratio of the isospin energy density to its zero temperature
Stefan-Bolzeman limit, the QCD phase diagram is investigated at a fixed 
low temperature, ${\cal T} = 20$ MeV for a range of isospin chemical potentials, 
from $\mu_I = m_\pi$ to $\mu_I = 4.5\ m_\pi$. An evidence of transition from
a pion gas to BEC is identified at $\mu_I = 1.30(7)\ m_{\pi}$.

\section{Acknowledgment}
The author thanks William Detmold and Kostas Originos for valuable discussions,
and is grateful for the support of the DOE NERSC facility, NSF XSEDE resources and 
in particular TG-PHY080039N, as well as the Sporades cluster at
the College of William \& Mary.

\end{document}